\documentclass[pss,fleqn,embeddedheads]{w-art}
\usepackage{w-thm}
\usepackage[]{graphicx}
\begin{document}
\DOIsuffix{theDOIsuffix}
\Volume{XX}
\Issue{1}
\Copyrightissue{01}
\Month{01}
\Year{2004}
\pagespan{1}{}
\Receiveddate{\sf zzz} \Reviseddate{\sf zzz} \Accepteddate{\sf
zzz} \Dateposted{\sf zzz}



\title[High fidelity measurement of singlet-triplet states in a quantum dot]{High fidelity measurement of singlet-triplet state in a quantum dot}


\author[T. Meunier]{T. Meunier\footnote{Corresponding
     author: e-mail: {\sf t.a.y.meunier@tudelft.nl}, Phone:+31\,15\,278\,6139
Fax: +31\,15\,278\,3251}\inst{1}}
\address[\inst{1}]{Kavli Institute of Nanoscience, Delft University of
Technology,\\
PO Box 5046, 2600 GA Delft, The Netherlands} 

\author[K-J. Tielrooij]{K-J. Tielrooij\inst{1}}

\author[I. T. Vink]{I. T. Vink\inst{1}}

\author[F. H. L. Koppens]{F. H. L. Koppens\inst{1}}

\author[H. P. Tranitz]{H. P. Tranitz\inst{2}}
\address[\inst{2}]{Institut f$\ddot{u}$r Angewandte und Experimentelle
Physik, Universit$\ddot{a}$t Regensburg, Regensburg, Germany}

\author[W. Wegscheider]{W. Wegscheider\inst{2}}

\author[L. P. Kouwenhoven]{L. P. Kouwenhoven\inst{1}}

\author[L. M. K. Vandersypen]{L. M. K. Vandersypen\inst{1}}


\begin{abstract}
We demonstrate experimentally a read-out method that distinguishes
between two-electron spin states in a quantum dot. This scheme
combines the advantages of the two existing mechanisms for
spin-to-charge conversion with single-shot charge detection: a large
difference in energy between the two states and a large difference
in tunnel rate between the states and a reservoir. As a result, a
spin measurement fidelity of $97\%$ was achieved, which is much
higher than previously reported fidelities.
\end{abstract}
\maketitle                   




\renewcommand{\leftmark}
{T. Meunier et al.: High fidelity measurement of singlet-triplet
state in a quantum dot}

Electron spins in quantum dots have received a lot of attention over
the last decade. The main reason is that the spin degree of freedom
is very well protected from the environment. Therefore, spins are
good candidates to be the building block of spintronic or quantum
information processing devices~\cite{LossDiVincenzo}. In this
context, it is desirable to be able to measure the spin state
associated to a single electron with high
fidelity~\cite{DiVicenzocrit}.

\begin{figure}[!t]
\center
\includegraphics[width=4in]{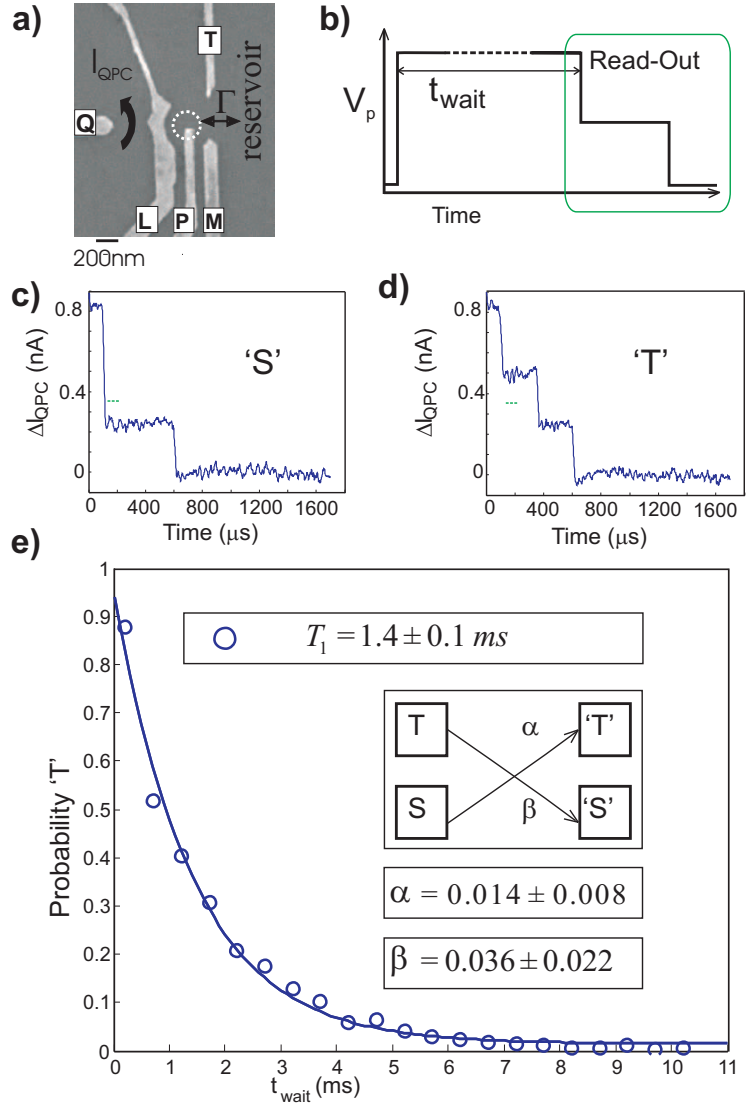}
\caption{(a) Scanning electronic micrograph showing the sample
design. (b) Voltage pulse applied to the gate 'P'. (c,d) Time
resolved QPC response when the two-electron state before the
measurement was declared singlet (c) and triplet (d) (the time
interval corresponds to the part surrounded by a square in the pulse
sequence presented in figure (b)). The threshold is shown with a
dashed line. (e) Probability of 'triplet' traces as a function of
the waiting time $t_{wait}$, out of 500 traces. The solid line is
the exponential fit to the data. $\alpha$ ($\beta$) is defined as
the probability for the measurement to return triplet 'T' (singlet
'S') if the actual state is singlet S (triplet T). $\alpha$ and
$\beta$ are obtained from the fit.} \label{Fig1}
\end{figure}

A direct spin state measurement is difficult because of the tiny
magnetic moment associated with the electron spin. In quantum dots,
single electron charge is easily measured. By correlating the spin
states to different charge states, it is possible to determine the
spin state in a single shot. Such spin to charge conversion has been
achieved experimentally in two ways. First, by positioning the
electrochemical potential of the lead in between the two relevant
spin states, one electron can tunnel off the dot from the
high-energy spin state whereas tunneling off the dot is
energetically forbidden from the ground
state~\cite{FujisawaNature,NatureReadout}. The fidelity of such
energy selective read-out (ERO), was limited to 82.5$\%$ by the
detector bandwidth~\cite{NatureReadout}. Indeed, in order to record
tunnel events, one has to make a compromise : tunneling has to be
fast enough to minimize relaxation before spin-to-charge conversion,
but not faster than the charge measurement bandwidth. In a second
method, one can distinguish two spins states when the rates for
tunneling on and off the dot are very different from each other. The
fidelity of such a tunnel rate selective read-out (TRRO) is limited
by the ratio of the two tunnel rates. Experimentally, a 90\%
fidelity was achieved, for a ratio of 20 in the two tunnel rates, in
agreement with theory.\cite{RonaldPRL}

Here, we propose to use an energy selective read-out assisted by a
difference in tunnel rates, for distinguishing between two-electron
singlet and triplet states in a single quantum dot. In this scheme,
an electron is energetically allowed to leave the dot if and only if
the spins are in a triplet state (which is here higher in energy
than the singlet state). The tunnel rate from the triplet state is
very fast. As a result, if the spins are in a triplet state, the
electron will almost always leave the dot before the spins have a
chance to relax to the singlet state. After one electron has left
the dot, another electron can tunnel into the dot again, into the
singlet state (the triplet is not energetically accessible). The
tunnel rate to the singlet state is very slow, much slower than the
charge measurement time in the experiment. The expected steps in the
charge detection signal (indicating that the dot contains
temporarily only one electron instead of two) will thus be clearly
visible. In summary, when the tunnel rate from the dot excited state
(triplet) to the reservoir is much larger than the tunnel rate from
the reservoir to the ground state of the dot (singlet), the two main
contributions to errors in the conventional energy selective
read-out are strongly suppressed (relaxation before tunneling, and
missed steps), and very high read-out fidelities could be achieved.

We test the high visibility energy read out with a quantum dot
(white dotted circle in Fig.~\ref{Fig1}(a)) and a quantum point
contact (QPC) defined in a two-dimensional electron gas (2DEG) with
an electron density of $1.3\cdot10^{15}$~m$^{-2}$, 90~nm below the
surface of a GaAs/AlGaAs heterostructure, by applying negative
voltages to gates $L$, $M$, $T$ and $Q$. Fast voltage pulses on gate
$P$ are used to rapidly change the electrochemical potential of the
dot. All measurements are performed at zero magnetic field. We tune
the dot to the few-electron regime~\cite{CiorgaPRB,JeroFewEl}, and
completely pinch off the tunnel barrier between gates $L$ and $T$,
so that the dot is only coupled to the reservoir on the
top~\cite{JeroAPL}. The conductance of the QPC is tuned to about
$e^2/h$, making it very sensitive to the number of electrons on the
dot~\cite{FieldPRL}. A voltage bias of 0.7~mV induces a current
through the QPC, $I_{QPC}$, of about 30~nA. Tunneling of an electron
on or off the dot gives steps in $I_{QPC}$ of 300~pA
\cite{LievenAPL,EnsslinAPL} and we observe them in the experiment
with a measurement bandwidth of 60~kHz, corresponding to a rise time
$t_R=5~\mu$s. The difference of tunneling rate between singlet and
triplet states arises from the distribution of electrons in the
orbitals for the two states. In the case of the singlet state, both
electrons are in the ground orbital whereas for the triplet state,
one electron is in the first excited orbital. The excited orbital
has a stronger overlap with the reservoir than the lowest orbital,
causing the tunnel rate to and from the triplet state, $\Gamma_T$,
to be much larger than the tunnel rate to and from the singlet
state, $\Gamma_S$~\cite{RonaldPRL}. In this measurement, when the
electrochemical of the reservoir is in between those associated to
singlet and triplet state $\Gamma_T/\Gamma_S\sim17$.

In order to extract the visibility, we reconstruct a relaxation
curve from the triplet to the singlet state. The protocol is
illustrated in Fig.\ref{Fig1}(b). The starting point is a dot with
one electron in the ground state. A first pulse is applied to gate
$'P'$ to move the singlet and the triplet electrochemical potentials
below the Fermi energy and a second electron tunnels into the dot.
In this situation, the ratio $\Gamma_T /\Gamma_S$ is higher than 17
and we observe that only the triplet state will be formed (perfect
initialisation in the excited state with an estimated error below
0.5\%). After a waiting time that we vary, we pulse the
electrochemical potential of the triplet state above the Fermi
energy while the electrochemical potential of the singlet is still
below. If the system is in the triplet state, an electron will
tunnel off the dot on a timescale $1/\Gamma_T\sim5~\mu$s (faster
than the measurement time resolution) and another electron will
tunnel on the dot to form a singlet on a timescale $1/\Gamma_S$
(slower than the measurement bandwidth and measured to be 7.8~kHz).
If the system is in the singlet state, tunneling is forbidden
energetically and the system remains in the singlet state (see
Fig.\ref{Fig1}(c,d)).

Due to the direct capacitive coupling of gate $P$ to the QPC
channel, $\Delta I_{QPC}$ follows the pulse shape (see Fig
\ref{Fig1}(c,d)). As a consequence of the tunneling events allowed
for triplet initial state, a step in the QPC response occurs during
the read-out stage. If $\Delta I_{QPC}$ goes above a predefined
threshold during the read-out stage then we conclude that the state
was triplet. If $\Delta I_{QPC}$ remains below the threshold we
conclude that the state was singlet. For each waiting time, we
record 500 individual traces and we extract the probability to
detect a triplet state. As expected, we observe an exponential decay
of the triplet population as a function of the waiting time, giving
a relaxation time, $T_1$, equal to 1.4$\pm$0.1~ms (see
Fig.\ref{Fig1}(e)). The experimentally determined measurement errors
are $\alpha=0.014$ and $\beta=0.036$, where $\alpha$ ($\beta$) is
defined as the probability for the measurement outcome to be triplet
(singlet) if the state was singlet (triplet).

$\alpha$ is mainly explained by thermally activated tunneling from
the singlet, a process suppressed in the experiment because the
energy splitting between the singlet and the reservoir, 0.45~meV, is
substantially larger than the electron temperature (20~$\mu$eV). Two
mechanisms are necessary to explain $\beta$. Some errors occur when
a triplet relaxed to a singlet before an electron tunnels off the
dot. The probability $\beta_1=1/(1+T_1\Gamma_T)$ of such a process
is $0.5\%$ in the present experimental set-up. The dominant error
process is tunneling into the dot on a timescale faster than the
charge measurement time. The probability of this error process is
$\beta_2=1-e^{-\Gamma_St_{R}}\sim4\%$.

We thus observe experimentally a read-out fidelity
$1-(\alpha+\beta)/2=97.5\%$ for distinguishing between the singlet
and the triplet states of a two-electron quantum dot. If the two
spin states would have the same tunneling rate, an optimal fidelity
equal to 93~$\%$ can be expected in the present measurement set-up
with an optimal tunnel rate of 17 kHz. We see thus that the
difference in tunnel rates between the two spin states significantly
improves the spin measurement fidelity. Alternatively, the fidelity
can be enhanced by increasing the bandwidth of the single electron
charge measurement, but this represents a real experimental
challenge.

\end{document}